\documentclass[twocolumn,aps,showpacs,prl]{revtex4}
\usepackage{amsmath}
\usepackage{graphicx}

\usepackage{dcolumn}
\usepackage{bm}
\usepackage{float}

\begin{document}

\title{Equilibration of Concentrated Hard Sphere Fluids}
\author{ Gabriel P\'erez-\'Angel$^1$, Luis Enrique S\'anchez-D\'iaz$^2$,
Pedro E. Ram\'irez-Gonz\'alez$^2$, Rigoberto Ju\'arez-Maldonado$^{2,3}$,
Alejandro Vizcarra-Rend\'on$^3$, and Magdaleno Medina-Noyola$^2$}

\address{(1)\ Departamento de F\'isica Aplicada CINVESTAV-IPN, Unidad M\'erida
Apartado Postal 73 Cordemex  97310. M\'erida, Yuc.,  Mexico}
\address{(2)\ Instituto de F\'{\i}sica {\sl ``Manuel Sandoval Vallarta"},
Universidad Aut\'{o}noma de San Luis Potos\'{\i}, \'{A}lvaro
Obreg\'{o}n 64, 78000 San Luis Potos\'{\i}, SLP, M\'{e}xico}
\address{(3)\ Unidad Acad\'emica de F\'isica, Universidad Aut\'onoma de Zacatecas,
Paseo la Bufa y Calzada Solidaridad, 98600, Zacatecas, Zac., Mexico}

\date{\today}

\begin{abstract}

We report a systematic molecular dynamics study of the isochoric
equilibration of hard-sphere fluids in their metastable regime close
to the glass transition. The thermalization process starts with the
system prepared in a non-equilibrium state with the desired final
volume fraction $\phi$ for which we can obtain a well-defined
\emph{non-equilibrium} static structure factor $S_0(k;\phi)$. The
evolution of the $\alpha$-relaxation time $\tau_\alpha(k)$ and
long-time self-diffusion coefficient $D_L$ as a function of the
evolution time $t_w$ is then monitored for an array of volume
fractions. For a given waiting time the plot of $\tau_\alpha(k;
\phi,t_w)$ as a function of $\phi$ exhibits two regimes
corresponding to samples that have fully equilibrated within this
waiting time $(\phi\le \phi^{(c)}(t_w))$, and to samples for which
equilibration is not yet complete $(\phi\ge \phi^{(c)}(t_w))$. The
crossover volume fraction $\phi^{(c)}(t_w)$ increases with $t_w$ but
seems to saturate to a value $\phi^{(a)} \equiv
\phi^{(c)}(t_w\to\infty) \approx 0.582$. We also find that the
waiting time $t_w^{eq}(\phi)$ required to equilibrate a system grows
faster than the corresponding equilibrium relaxation time,
$t_w^{eq}(\phi) \approx 0.27
\times[\tau_{\alpha}^{eq}(k;\phi)]^{1.43}$, and that both
characteristic times increase strongly as $\phi$ approaches
$\phi^{(a)}$, thus suggesting that the measurement of
\emph{equilibrium} properties at and above $\phi^{(a)}$ is
experimentally impossible.

\end{abstract}


\maketitle

Above a certain size polydispersity, real and simulated hard sphere
liquids fail to crystalize for volume fractions $\phi$ beyond the
freezing point $\phi^{(f)} = 0.494$ of the monodisperse system
\cite{alder,pusey1,zaccarelli,pusey2}. As $\phi$ increases the
viscosity increases enormously, and the metastable liquid eventually
becomes an amorphous solid. Mode coupling theory (MCT)
\cite{goetze1} predicts a transition from metastable fluid to ideal
nonergodic states, characterized by the vanishing of the long-time
self-diffusion coefficient $D_L$ and the divergence of both, the
$\alpha$-relaxation time $\tau_\alpha$ and the viscosity $\eta$. For
the hard-sphere fluid the phenomenology predicted by MCT at
$\phi^{(a)} \approx 0.52$ has been essentially confirmed by the
experimental observations in hard-sphere colloidal suspensions at
$\phi^{(a)}_{exp} \approx 0.58$ \cite{vanmegen1,vanmegen2}, although
a number of intrinsic experimental uncertainties render the precise
determination of $\phi_{exp}^{(a)}$ a topic of recurrent scientific
discussion
\cite{vanmegen1,vanmegen2,pusey2,segre1,brambilla,elmasri}.

The recent work of Brambilla et al. \cite{brambilla,elmasri},
however, seems to put the very experimental relevance of the
divergent scenario predicted by MCT under severe questioning. By
fitting their dynamic light scattering data with the asymptotic
expression $\tau_\alpha (\phi) \sim (\phi^{(a)}-\phi)^{-\gamma}$,
traditionally associated with MCT, these authors determined
$\phi^{(a)}$ to be $\phi^{(a)} = 0.590 \ \pm \ 0.005$. If the ideal
MCT picture were to be observed in their experiments, the measured
$\tau_\alpha (\phi)$ should be infinite for $\phi > \phi^{(a)} $.
Instead, for the volume fraction range $\phi^{(a)} < \phi \lesssim
0.6$, they report large but finite relaxation times, determined
through an extremely careful experimental procedure designed to deal
with artifacts caused, for example, by sample heating or
sedimentation, which allowed them to accurately monitor the
equilibration process of their samples \cite{elmasri}. Thus, the
most immediate interpretation is that these measurements involve
macroscopic states in which the system, instead of falling out of
equilibrium, remains ergodic and enters a new dynamical regime where
$\tau_\alpha$ increases with volume fraction according to a
different functional form, namely, $\tau_\alpha (\phi) \sim
\tau_\infty \exp[A(\phi_0-\phi)^{-\delta}]$.

This interpretation, however, rests on the assumption that all
measurements reporting an apparent stationary behavior indeed
involve fully equilibrated systems. Recent molecular dynamics
simulations \cite{kimsaito,Hermes}, however, suggest that this 
assumption should not be taken for granted without further discussion. 
For example, according to \cite{kimsaito},  the relaxation
time $\tau_{\mathrm{hetero}}$ of dynamic heterogeneities may grow
like $\tau_{\mathrm{hetero}} \sim \tau_\alpha^{1.5}$ as the glass
transition is approached. Thus, if one has to wait until ``slow"
regions become ``fast" regions and vice versa, one possibility that
cannot be ruled out is that when the
\emph{equilibrium} relaxation time $\tau^{eq}_\alpha(\phi)$ indeed
diverges, the system will require a similarly divergent
\emph{equilibration} time $t^{eq}_w(\phi)$, thus blurring even the
most accurate observation. Motivated in part by these
considerations, here we \emph{intentionally} study the effects on  
$\tau_\alpha(\phi)$ of the \emph{incomplete} equilibration of 
concentrated hard-sphere systems close to the glass transition by 
means of systematic computer simulations, in which some of the 
intrinsic uncertainties of the experimental samples will be absent.

As in \cite{Hermes}, the basic simulation experiment consists of monitoring the
irreversible evolution of a hard-sphere system initially prepared at
a non-equilibrium state characterized by a prescribed volume
fraction $\phi$ and by a well-defined non-equilibrium static structure factor
$S_{0}(k;\phi)$. The irreversible evolution to equilibrium is then
described in terms of the time-evolving non-equilibrium static
structure factor $S_{t_w}(k;\phi)$ and self intermediate scattering
function (Self-ISF) $F_S(k,\tau,t_w)$, where $t_w$ is the evolution
(or ``waiting") time after the system was prepared. The naturally expected long-$t_w$
asymptotic limit of these properties is, of course, the
\emph{equilibrium} static structure factor $S^{eq}(k;\phi)$ and
self-ISF $F_S^{eq}(k,\tau)$. Our interest is to determine the volume
fractions for which equilibrium is reached within a given waiting
time $t_w$.

We use event-driven molecular dynamics to simulate the evolution of
$N = 1000$ particles in a volume $V$, with particle diameters
$\sigma$ evenly distributed between $\overline{ \sigma} (1-w/2)$ and
$\overline{ \sigma} (1+w/2)$, with $\overline \sigma$ being the mean
diameter. We consider the case $w=0.3$, corresponding to a
polydispersity $s_\sigma = w/\sqrt{12}=0.0866$. According to the
results reported in \cite{zaccarelli}, at this
polydispersity the system shows no evidence of crystallization for
any volume fraction $\phi=(\pi/6) n \overline{\sigma^3}$, where
$\overline{\sigma^3}$ is the third moment of the size distribution
and $n$ is the total number density $n\equiv N/V$. All the particles
are assumed to have the same mass $M$. The length, mass, and time
units employed are, respectively, $\overline{\sigma}$, $M$, and
$\overline{\sigma}\sqrt{M/k_BT}$.

\begin{figure}
\begin{center}
\includegraphics[scale=.18]{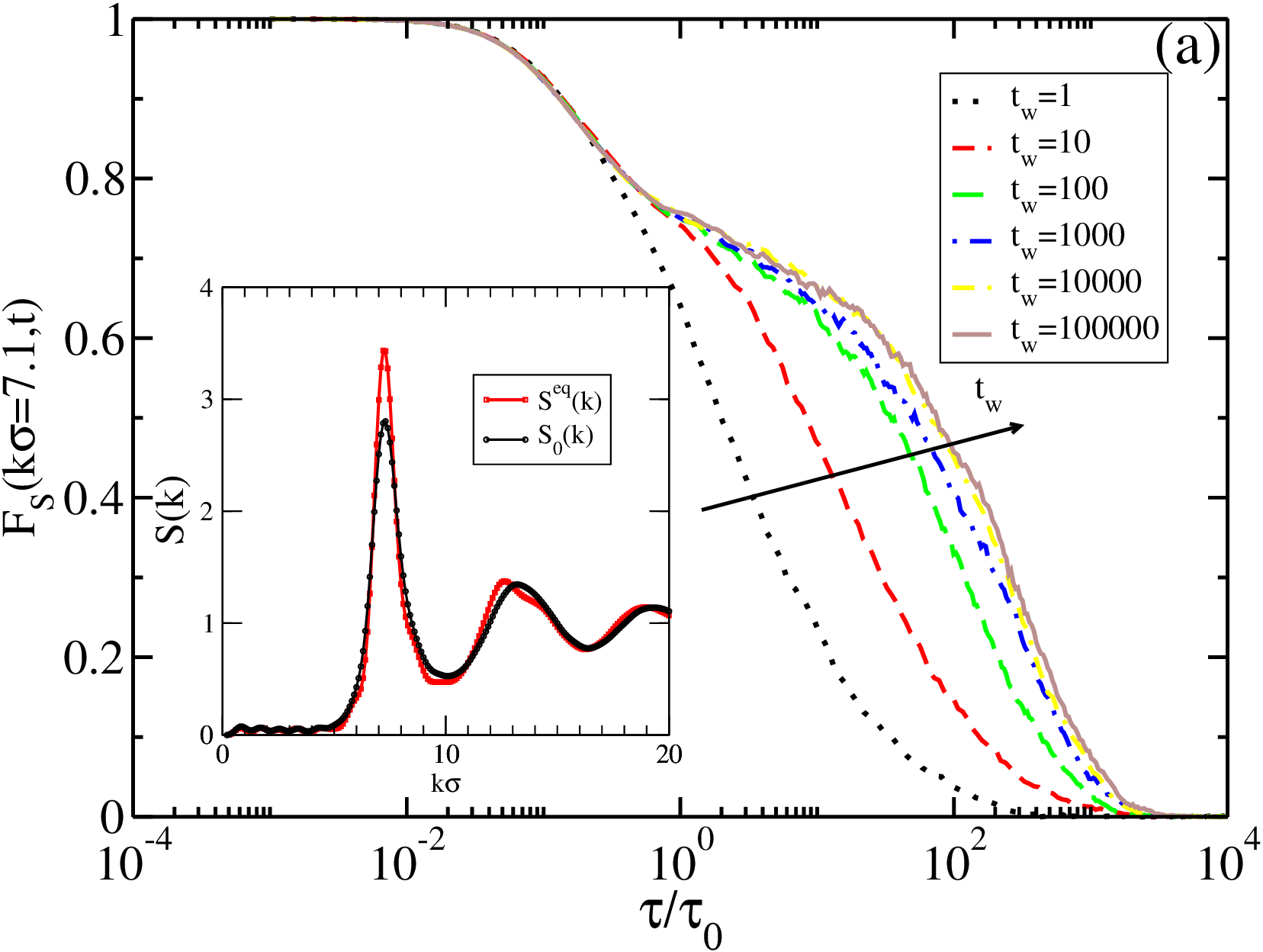}
\includegraphics[scale=.18]{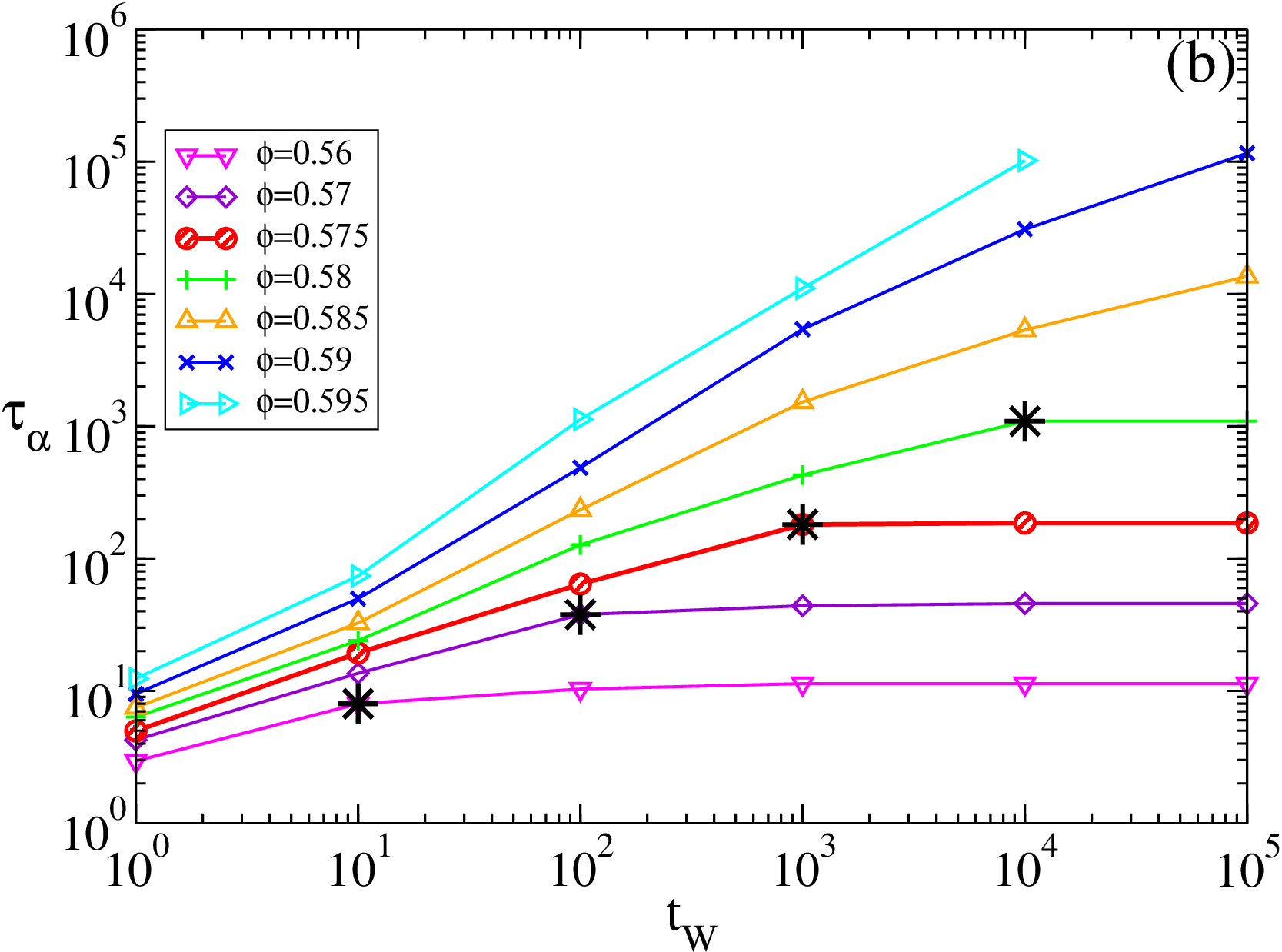}
\caption{(a)Self intermediate scattering function $F_S(k,\tau;t_w)$ of a 
polydisperse hard-sphere system ($s=0.0866$) 
evaluated at $k = 7.1$  at volume fraction $\phi=0.575$ and
polydispersity $s=0.0866$ as a function of the correlation time
$\tau$ for waiting times $t_w=10^0, \ 10^1,...,10^5$. The inset of
(a) shows the corresponding $S_{0}(k;\phi)$ (black circles) and
$S^{eq}(k;\phi)$ (red squares).(b)Simulation data of the 
$\alpha$-relaxation time $\tau_{\alpha}(k;\phi,t_w)$ as a function 
of $t_w$ at fixed volume fraction.The asterisks highlight 
the points $(t_w^{eq}(\phi),\tau_{\alpha}^{eq}(k;\phi))$. } \label{fig1}
\end{center}
\end{figure}

To produce the initial configurations we used soft-particle
molecular dynamics to simulate the evolution of a set of initially
overlapping, randomly placed particles, with the correct
distribution of diameters, interacting through a short-ranged
repulsive soft (but increasingly harder) interaction, and in the
presence of strong dissipation. For $\phi$ below the random close
packing limit, this system evolves rapidly into a disordered
configuration with no overlaps. These non-thermalized hard-sphere
configurations are then given random velocities generated by a
Maxwell-Boltzmann distribution, with $k_B T$ set as the energy unit.
These configurations are then used as the starting configurations
for the event-driven simulation of the HS equilibration process.

The simulations were carried for an array of values of $\phi$
between $0.480$ and $0.595$. For each such volume fraction we used
waiting times from $1$ to $10^5$ in powers of $10$. The sequence of
configurations obtained is employed to generate the Self-ISF
$F_S(k,\tau,t_w)\equiv (1/N)\langle\sum_{i=1}^N \exp {[i{\bf
k}\cdot({\bf r}_i(t_w+\tau)-{\bf r}_i(t_w))]}\rangle$ and the mean
squared displacement (MSD) $\langle(\Delta
\textbf{r}(\tau;t_w))^2\rangle \equiv (1/N)\langle\sum_{i=1}^N [{\bf
r}_i(t_w+\tau)-{\bf r}_i(t_w)]^2\rangle$, where ${\bf r}_i(t)$ is
the position of the $i$th particle at time $t$, $\tau$ is the
\emph{correlation} time, and the brackets indicate averaging over
(at least) $20$ independent realizations. $F_S(k,\tau;t_w)$ is
evaluated at $k = 7.1$, close to the main peak of $S^{eq}(k;\phi)$
for all the values of $\phi$ considered. The $\alpha$-relaxation
time $\tau_{\alpha}(k;\phi,t_w)$ is defined by the condition
$F_S(k,\tau_{\alpha},t_w) = 1/e$, and the long-time self-diffusion
coefficient $D_L$ by $D_L (\phi;t_w)\equiv \lim_{\tau \to \infty}
\langle(\Delta \textbf{r}(\tau;t_w))^2\rangle / 6\tau $.

Let us illustrate the results of this procedure for one specific
volume fraction, namely, $\phi=0.575$. In Fig. \ref{fig1}(a) we 
present the simulation results for $F_S(k,\tau;t_w)$ evaluated at 
$k = 7.1$ as a function of the correlation time $\tau$ for the 
sequence of waiting times $t_w=10^0, \ 10^1,...,10^5$. This 
sequence exhibits the increasing slowing down of the dynamics as 
the system approaches its equilibrium state and illustrates the 
fact that $F_S(k,\tau;t_w)$ saturates to its equilibrium value
$F_S^{eq}(k,\tau)$ after a certain \emph{equilibration waiting time}
$t_w^{eq}(\phi)$. For example, from the illustrative data in the
figure we find that $t_w^{eq}(\phi=0.575) \approx 10^4$. A similar
evolution and saturation is observed in the static structure factor
$S_{t_w}(k;\phi)$, which exhibits, as expected, a large increase at
the first diffraction peak. The inset of Fig. \ref{fig1}(a) presents
the initial static structure factor $S_{0}(k;\phi)\equiv
S_{t_w=0}(k;\phi)$ and the final equilibrium $S^{eq}(k;\phi)$. From
the data for $F_S(k,\tau;t_w)$ in this figure we can determine the
$\alpha$-relaxation time $\tau_{\alpha}(k;\phi,t_w)$ as a function
of $t_w$. The results indicate that the $\alpha$-relaxation time
$\tau_{\alpha}(k;\phi=0.575,t_w)$ saturates approximately to its
equilibrium value $\tau_{\alpha}^{eq}(k;\phi=0.575) \approx 2 \times
10^2$ within the equilibration waiting time $t_w^{eq}(\phi=0.575)
\approx 10^4$.

Fig. \ref{fig1}(b) plots the dependence of the $\alpha$-relaxation
time $\tau_{\alpha}(k;\phi,t_w)$ as a function of waiting time $t_w$
for fixed volume fraction $\phi$. These plots confirm that beyond an
equilibration waiting time $t_w^{eq}(\phi)$, the $\alpha$-relaxation
time $\tau_{\alpha}(k;\phi)$ saturates approximately to its
equilibrium value $\tau_{\alpha}^{eq}(k;\phi)$. To emphasize these
concepts we have highlighted the points $(t_w^{eq}(\phi),
\tau_{\alpha}^{eq}(k;\phi))$ in the figure. In fact, we notice that
the highlighted points $(t_w^{eq}(\phi),
\tau_{\alpha}^{eq}(k;\phi))$ obey the approximate relation
$t_w^{eq}(\phi) \approx 0.27
\times[\tau_{\alpha}^{eq}(k;\phi)]^{1.43}$, suggesting that the
waiting time $t_w^{eq}(\phi)$ required to equilibrate a system is
always longer than the corresponding equilibrium relaxation time
$\tau_{\alpha}^{eq}(k;\phi)$, and that both characteristic times
increase strongly with $\phi$.

\begin{figure}[H]
\begin{center}
\includegraphics[scale=.25]{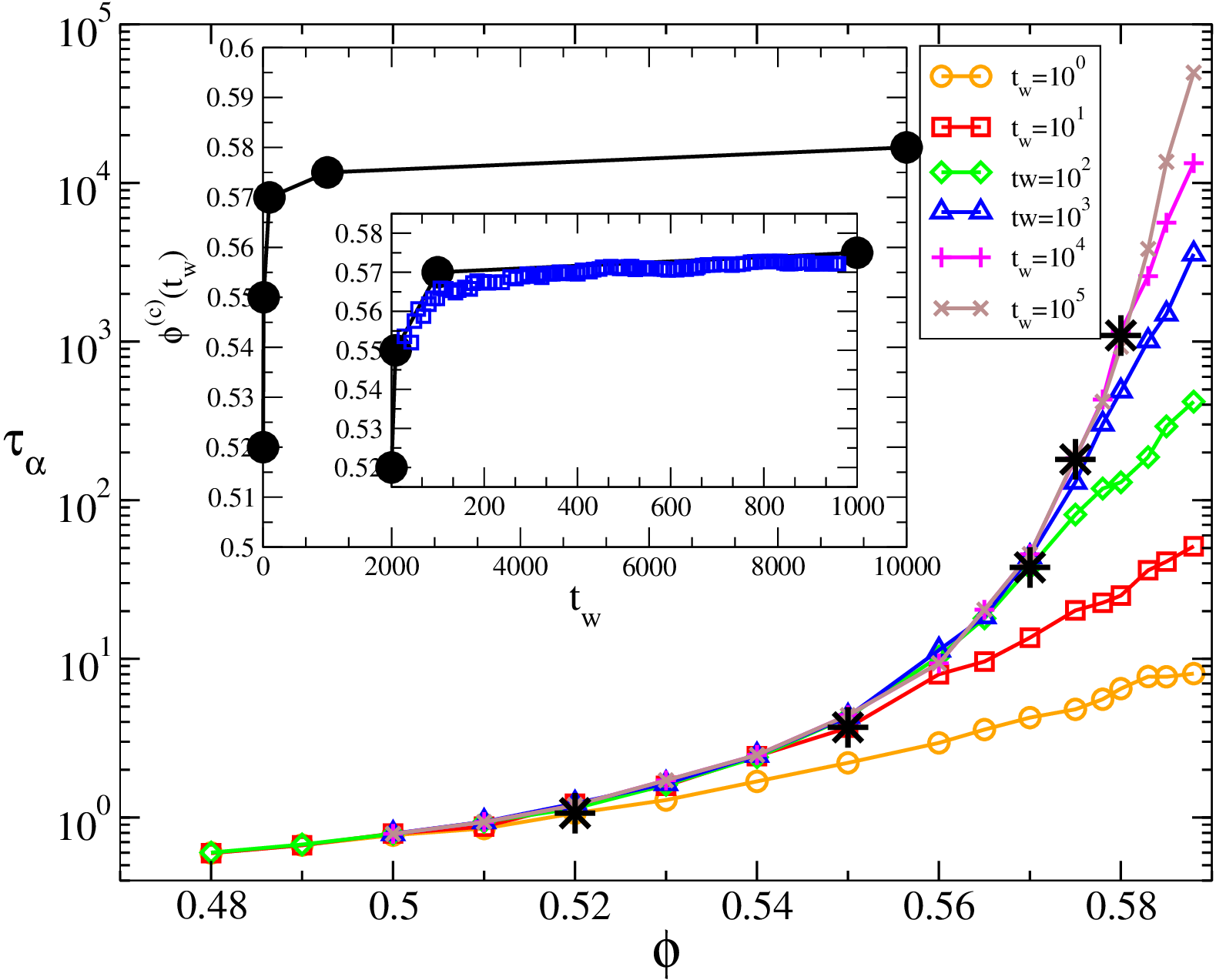}
\caption{Simulation data of the $\alpha$-relaxation time
$\tau_{\alpha}(k;\phi,t_w)$ of a polydisperse hard-sphere system
($s=0.0866$) as a function of volume fraction  at fixed waiting time. 
The asterisks indicate, in this case, the crossover volume 
fraction $\phi^{(c)}(t_w)$ at the various waiting times. The circles in the 
inset display the evolution of $\phi^{(c)}(t_w)$ with 
waiting time, and 
the squares are the results in Fig. 5(a) of  Ref. \cite{Hermes}.} \label{fig2}
\end{center}
\end{figure}

Just like Hermes and Dijkstra \cite {Hermes} did with the pressure (see their Fig. 1), 
let us display our data for $\tau_{\alpha}(k;\phi,t_w)$ of Fig.
\ref{fig1}(b) in a complementary manner,
namely, as a function of
volume fraction for fixed waiting time $t_w$, and this is done in
Fig. \ref{fig2}. The first feature to notice in each of the
corresponding curves is that one can distinguish two regimes in
volume fraction, namely, the low-$\phi$ (equilibrated) regime and
the high-$\phi$ (non-equilibrated) regime, separated by a crossover
volume fraction $\phi^{(c)}(t_w)$. Focusing, for example, on the
results corresponding to $t_w=10^3$, we notice that
$\phi^{(c)}(t_w=10^3)\approx 0.57$. In Fig. \ref{fig2} we have
highlighted the crossover points $(\phi^{(c)}(t_w),
\tau^{eq}_{\alpha}(k;\phi))$. We observe that the resulting
crossover volume fraction $\phi^{(c)}(t_w)$ first increases rather
fast with $t_w$, but then slows down considerably, suggesting a
saturation to a value slightly larger than 0.58, as indicated in the
inset of Fig. \ref{fig2}, which also include  the results for  
$\phi^{(c)}(t_w)$ determined by Hermes and Dijkstra \cite{Hermes} from 
the pressure, denoted by $\eta_g$ in their Fig. 5(a). The estimate of 
the limit $\phi^{(c)}(t_w\to \infty)$ will hardly be determined by even 
more powerful simulations, and the need of a theoretical framework is 
required. 

One of the main products of the simulation results just presented 
is the determination of the volume fraction dependence of the
\emph{equilibrium} $\alpha$-relaxation time
$\tau_{\alpha}^{eq}(k;\phi)$. Clearly, our simulation experiment can
determine this property only within the window $0\le \phi \le
\phi^{(c)}(t_w^{max})$, where $t_w^{max}$ is the maximum waiting
time achieved in the simulation experiment. In our case, $t_w^{max}
=10^5$ yielding $\phi^{(c)}(t_w^{max})\approx 0.58$. These results,
scaled with $\tau_{\alpha,0}(k;\phi)\equiv 1/ k^2 D^0$, with
$D^0=\sqrt{\pi}/16\phi$ (see below), are plotted in Fig.
\ref{angell-tau-eq} as solid squares. For $ \phi \ge 0.58$ the
$t_w$-dependent $\alpha$-relaxation time $\tau_{\alpha}(k;\phi,t_w)$
did not saturate to its equilibrium value within the total duration
of the present simulation experiment. These results are also plotted
in Fig. \ref{angell-tau-eq} as empty squares, to denote insufficient
equilibration. Thus, only the data in solid squares are meaningful
when comparing with the predictions of \emph{equilibrium} theories
such as MCT or the more recently developed self-consistent
generalized Langevin equation (SCGLE) theory \cite{todos2}.

MCT and the SCGLE theory provide similar answers regarding the
asymptotic divergence of the relaxation times. We consider, however,
that there is no need to appeal to asymptotic expressions, which
have a more restricted range of validity, when one has easy access
to the full numerical solution of the corresponding theory, as we do
for the SCGLE theory of colloid dynamics. As we have recently
discovered \cite{atomicvscoloidal}, the latter theory also describes
the long-time dynamics  of atomic systems provided the solvent
short-time self-diffusion coefficient $D^0$ is replaced by the
kinetic-theory self-diffusion coefficient, given by
$D^0=\sqrt{\pi}/16\phi[\sigma\sqrt{k_BT/M}]$
\cite{mcquarrie,chapmancowling}. In Fig. \ref{angell-tau-eq} we
compare the simulated equilibrium data for $\tau^*(k;\phi,t_w)\equiv
k^2D^0 \tau_{\alpha}(k;\phi,t_w)$  ($ \phi \le 0.58$)  with the
predictions of the SCGLE theory (Ecs. (1), (2), and (5-8) of Ref
\cite{todos2}, with $k_c=8.48$ adjusted to fine-tune the comparison 
with these equilibrium data). According to this fit,
$\tau_{\alpha}^{eq}(k;\phi)$ would diverge at $\phi^{(a)} \approx
0.582$.

There is, of course, no reason to include the non-equilibrated data
of Fig. \ref{angell-tau-eq} in this comparison. As a mere fitting
exercise, however, we notice that the full set including these
non-equilibrated data can be fitted by the expression $\tau_\alpha
(\phi) = \tau_\infty \exp[A(\phi_0-\phi)^{-\delta}]$, thus finding
$A=0.02$, $\delta=1.921$,  $\tau_\infty=0.21$ and $\phi_0 = 0.6235$
(dashed line in the figure). Amazingly enough, we find that this
functional form provides a reasonable fit also for the shorter
waiting times $t_w= \ 10^4$ and $10^3$ using the same values for
$\delta$, $C$, and $ \tau_\infty$, but with $\phi_0$ = 0.631 and
0.635, respectively, as illustrated in the inset (a) of Fig. 
\ref{angell-tau-eq}.

In order to relate our simulation results with the experimental
observations of Refs. \cite{brambilla, elmasri}, in the inset (b)
of Fig. \ref{angell-tau-eq} we compare the experimental
equilibration data of Fig. 6 of Ref. \cite{elmasri} for the sample
labeled $\phi^{exp}= 0.5876 $, with the simulation data
corresponding to $\phi= 0.58 $ in Fig. \ref{fig1}(b) above. The
excellent agreement between the simulated and the experimental
equilibration data suggests that the difference in the value of
$\phi$ and $\phi^{exp}$ could be explained by the intrinsic
uncertainties discussed in Ref. \cite{elmasri} regarding the
determination of the volume fraction of the system.

Assuming that this is the case, we can directly compare our MD
simulation results in Fig.\ref{angell-tau-eq} for
$\tau_{\alpha}^{eq}(k;\phi)/\tau_{\alpha,0}^{eq}(k;\phi)$ with the
experimental data in Fig. 13 of Ref. \cite{elmasri}, provided that
we assume a constant ratio $\phi/\phi^{exp}$. The dark circles in
that figure are precisely those experimental data as a function of
the experimental volume fraction reduced by a factor
$\phi/\phi^{exp}$ (fitted to yield the value 0.985), to
approximately account for the referred uncertainties. In addition,
as a simple manner to treat hydrodynamic interactions \cite{prlhi},
we have to take into account that the role of the normalizing
parameter $D^0$ is played, in the experimental data, by the
short-time self-diffusion coefficient $D_S(\phi)$, given
approximately by $D_S(\phi)/D_S(\phi=0)= (1-\phi)/(1+1.5\phi)$
\cite{mazurgeigen}. This comparison suggests a completely similar
phenomenology, although it is quite clear only in the case of our
simulation data that the departure of
$\tau_{\alpha}(k;\phi,t_w=10^5)$ from the equilibrium curve
predicted by the SCGLE theory near $\phi^{(a)}$ is due to the
insufficient equilibration of the system within the maximum waiting
time $t^{max}_w=10^5$ of our simulation experiment.

\begin{figure}
\begin{center}
\includegraphics[scale=.25]{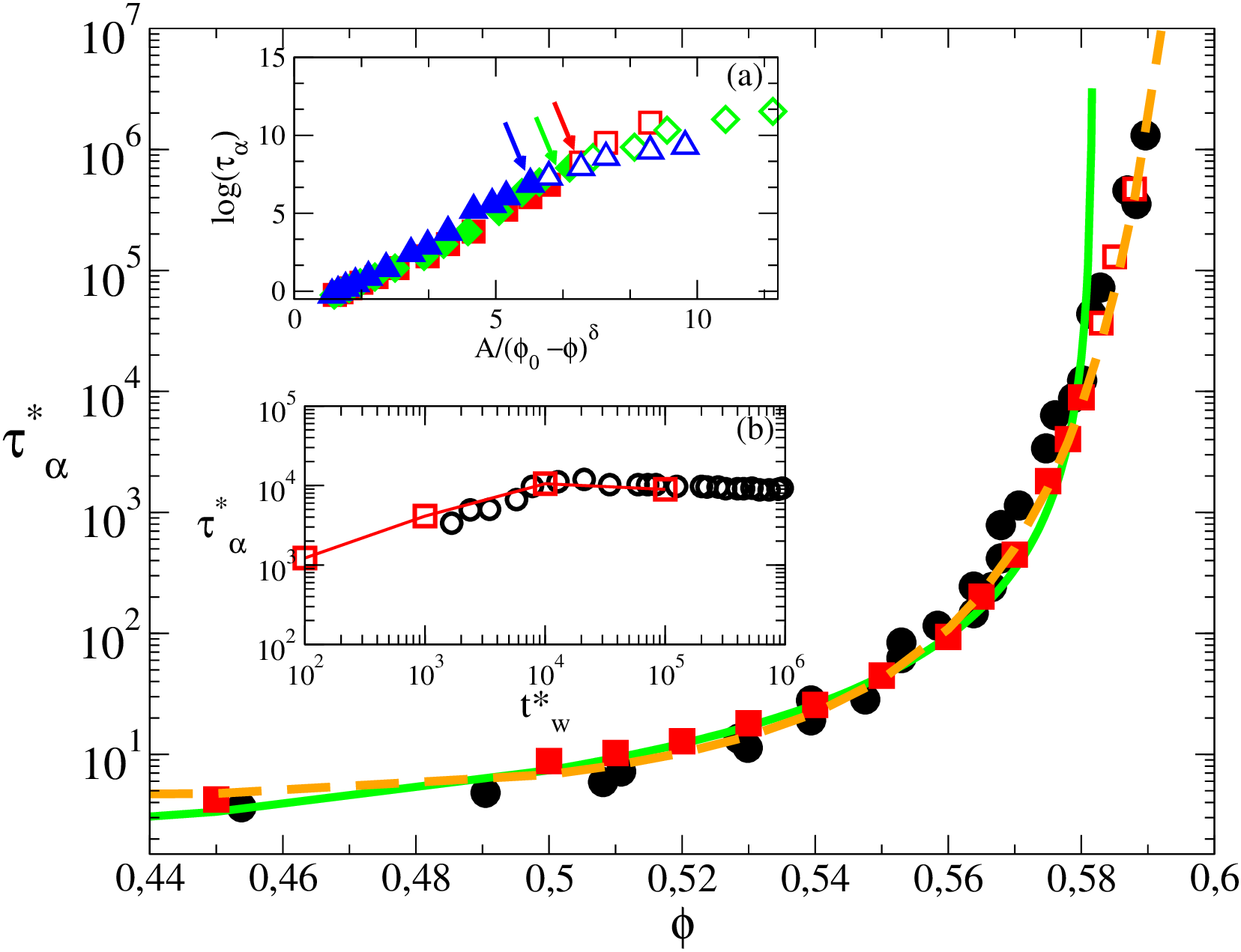}
\caption{Volume fraction dependence of the scaled
$\alpha$-relaxation time $\tau^*(k;\phi,t_w)\equiv k^2D^0
\tau_{\alpha}(k;\phi,t_w)$. The solid (empty) squares denote
simulation data of fully equilibrated (insufficiently equilibrated)
systems. The solid line are the predictions of the SCGLE theory.
The dashed line is the fit with $\tau_\alpha (\phi) =
\tau_\infty \exp[A(\phi_0-\phi)^{-\delta}]$. The solid circles 
correspond to the experimental results of Fig. 13 of Ref.
\cite{elmasri}. The inset (a) plots $\tau_{\alpha}(k;\phi,t_w)$ 
as a function of $A(\phi_0-\phi)^{-\delta}$ for $t_w= 10^5$ 
(squares), $\ 10^4$ (diamonds) and $10^3$ (triangles), with the 
arrows pointing at the corresponding crossover volume fraction 
$\phi^{(c)}(t_w)$. The inset (b) compares our simulation results 
(empty squares with line) for $\tau^*(k;\phi,t_w)$ vs. 
$t_w^*\equiv k^2 D^0 t_w$ for $\phi= 0.58$ with the experimental 
data of Fig. 6 of Ref. \cite{elmasri} (empty circles) at 
$\phi^{exp}= 0.5876 $.} \label{angell-tau-eq}
\end{center}
\end{figure}

The results just presented suggest, however, that any simulation
aimed at determining the \emph{equilibrium} value of dynamic order
parameters such as $\tau_{\alpha}(k;\phi,t_w)$ and $D_L (\phi;t_w)$
near the dynamic arrest transition is bound to be limited by the
duration of the simulation experiment, represented by the maximum
waiting time $t_w$ involved. This limits the determination of these
equilibrium values to the window of volume fractions $0\le \phi \le
\phi^{(c)}(t_w^{max})$. For $\phi \ge \phi^{(c)}(t_w^{max})$, the
simulation results will be reporting the properties of an
insufficiently equilibrated system. The results presented here
indicate that if we want to enlarge this window we would have to go
to exponentially longer waiting times, which is bound sooner or
later to become a lost battle. There is, of course, no obvious
reason to believe that a different situation will prevail in
experimental samples.

Let us finally notice that the expression $\tau_\alpha (\phi) =
\tau_\infty \exp[A(\phi_0-\phi)^{-\delta}]$ gives a reasonble
fit for our non-equilibrium data, even at early stages in the
waiting time $t_w$, suggesting that this dynamical regime could
be consecuence of the lack of equilibration. These means that
the correct analysis of the data corresponding to incompletely
equilibrated conditions must be made in the framework of a
non-equilibrium theory. It is pertinent to mention that our original motivation to carry
out the present simulations was precisely the need to generate
reliable data  incompletely equilibrated systems that will serve
as a reference to test the recently developed non-equilibrium extension \cite{nonEqSCGLE} of the  SCGLE theory.

ACKNOWLEDGMENTS: This work was supported by the Consejo Nacional de
Ciencia y Tecnolog\'{\i}a (CONACYT, M\'{e}xico), through grant No.
84076 and CB-2006-C01-60064, and by Fondo Mixto CONACyT-SLP through
grant FMSLP-2008-C02-107543.

\end{document}